# ε-differential agreement: A Parallel Data Sorting Mechanism for Distributed Information Processing System

Wei Bi*, Xiangyu Liu, Maolin Zheng

Seele Tech Corporation, San Francisco, USA
weibi@seelenet.com

**Abstract:** The order of the input information plays a very important role in a distributed information processing system (DIPS). This paper proposes a novel data sorting mechanism named the ε-differential agreement (EDA) that can support parallel data sorting. EDA adopts the collaborative consensus mechanism which is different from the traditional consensus mechanisms using the competition mechanism, such as PoS, PoW, etc. In the system that employs the EDA mechanism, all participants work together to compute the order of the input information by using statistical and probability principles on a proportion of participants. Preliminary results show variable fault-tolerant rates and consensus delay for systems that have different configurations when reaching consensus, thus it suggests that it is possible to use EDA in a system and customize these parameters based on different requirements. With the unique mechanism, EDA can be used in DIPS of multi-center decision cluster, not just the rotating center decision cluster.

## 1 Introduction

The sequential consistency is the most important thing to record actions[1], especially in DIPS used by financial and trading platforms. Any little change in data sequence will lead to simultaneous failure of multiple transactions or even destroys the whole system. In a traditional centralized system, the order of data is determined by the central database. With the increasing number of terminals, instantaneous requirements put a huge strain on the system. In general, it costs a lot for data service providers to maintain the stability and security of the central database. Small companies can't afford it and the only choice is "face the fate". If the central database is damaged, the whole system usually rolls back to the last backup.

As a completely different thinking from a traditional system, the blockchain network adopts mutual consensus and verification to record actions, which solves the problem that a central database is facing. In a blockchain network, the order of data is determined by the peer that obtains the right to build the block whom we called the winner miner. Every peer in the blockchain network is a candidate, and each peer may want to record different data. What the network doing is to select one lucky dog from all candidates in a period, and others must record a replica of the data of what the lucky one says, that is the rule. This method provides a stable and non-changeable data recording ability, but this serial consensus is too slow, and cannot provide adequate performance in real world scenario.

In this paper, we propose the EDA mechanism which is original inspired by how neurons in the brain work. The EDA mechanism is designed to provide an effective and practical user-friendly parallel data sorting mechanism for application scenarios in DIPS.

## 2  System description

The DIPS can be simply defined as 3 conditions:
1. A set of peers $\wp$. The number of peers is denoted as $|\wp|$. In the case of a public system, $|\wp|$ is unknown;

2. A database on each peer $p_i \in \wp$, which consists of ordered transactions $\{c^{s_i}\}$ with local ordering $s_i$;

3. A consensus protocol achieves agreement over the ordering of transactions and ensures that $\forall p_i, \forall p_j: c^{s_i} = c^{s_j} \rightarrow |s_i - s_j| \leq \varepsilon$, which means that the same transaction has the same ordering on any pair of nodes within error of $\varepsilon$. Deterministic protocols such as PoW and PBFT is a special case with $\varepsilon = 0$.

With the conditions above, the complex system can be parametrized for modeling and analysis.

## 3  Principle

The generation function $g(x)$ and the activation function $a(x)$ are used to describe how the nervous works [2, 3], and we assume that the whole EDA network is a cluster of neurons. For one transaction $c$, at the beginning of the consensus at round $t = 0$, the order of the transaction is denoted as $s_i^0$ which is randomly broadcasted to a random $p\%$ proportion of other nodes. The consensus then proceeds in consecutive rounds $t > 0$ at each node $s_i^t = a\left(g(\mathbf{s}_{\rightarrow i}^{t-1})\right)$ where $\mathbf{s}_{\rightarrow i}^{t-1}$ denotes the ordering received at node $i$ in round $t - 1$.

Each peer is acting as a neuron using $g(x) = median(x)$ and $a(g(x)) = g(x)$. We also assume that $x$ is ordered with $M$ elements. It is well known that the probability of $100p (p \in [0,1])$ percentile, namely $P_{100p}$, falling between $x_k$ and $x_{k+1}$ follows a

binomial distribution
$$Pr(x_k \leq P_{100p} \leq x_{k+1}) = C_M^k p^k (1-p)^{M-k}$$
$M$ is the dimension of $x$. This distribution can be approximated by a normal distribution
$$Pr(x_k \leq P_{100p} \leq x_{k+1}) \approx N(Mp, Mp(1-p))$$

when the $Mp(1-p) > 9$. For median $p = 0.5$, the order statistics of $s_i^t$ become
$$o(s_i^t) = median(s_{\rightarrow i}^{t-1}) \sim N(0.5 M_i^{t-1}, 0.25 M_i^{t-1})$$

where $o(\cdot)$ is the order statistics and $M_i^{t-1}$ is the number of ordering received from last round at node $i$. This means that the order $s_i^t$ is at the $\lfloor 0.5 M_i^{t-1} \rfloor th$ element in $s_{\rightarrow i}^{t-1}$ with the standard deviation of $0.5\sqrt{M_i^{t-1}}$. By dividing the total number $M_i^{t-1}$, the order statistic can be re-expressed in percentiles as
$$Pr(o(s_i^t)) = N\left(0.5, \frac{0.25}{M_i^{t-1}}\right)$$

with respect to $Pr(s_i^{t-1})$, showing that larger sample size of $s_{\rightarrow i}^{t-1}$ gives rise to tighter confidence interval. This is in comparison with the standard deviation of $Pr(s_i^{t-1})$ that is 0.3413 away from the mean according to the definition of a Normal distribution. When $M_i^{t-1} > 2$, $Pr(o(s_i^t))$ reduces the standard deviation to $\frac{0.5}{\sqrt{M_i^{t-1}}}$.

The ordering received by all nodes $\cup_i s_{\rightarrow i}^{t-1}$ forms a complete set of estimates of ordering made by all peers. Assuming that $s_{\rightarrow i}^{t-1}$ are random sample from this complete set with the same underlying distribution so $Pr(o(s_i^{t-1}))$ is identical for all nodes. The distribution of the difference $o(s_i^t) - o(s_j^t)$ between two arbitrary different peers follows a Normal distribution
$$Pr(o(s_i^t) - o(s_j^t)) = N\left(0, \frac{0.25}{M_i^{t-1}} + \frac{0.25}{M_j^{t-1}}\right)$$

To simplify the derivation, we assume that $M_i^{t-1} = M_j^{t-1} = M^{t-1}$ so the difference between two peers follows the distribution
$$Pr(o(s_i^t) - o(s_j^t)) = N\left(0, \frac{0.5}{M^{t-1}}\right)$$

with respect to the distribution in the last round $Pr(o(s_i^{t-1})) = Pr(o(s_j^{t-1}))$ which are also normally distributed. Note that the standard deviation of $Pr(o(s_i^{t-1}))$ is between the 0.1587 and 0.8413 in percentile by the definition of a Normal distribution. The distribution change during the EDA working can be seen in Fig. 1 and Fig. 2

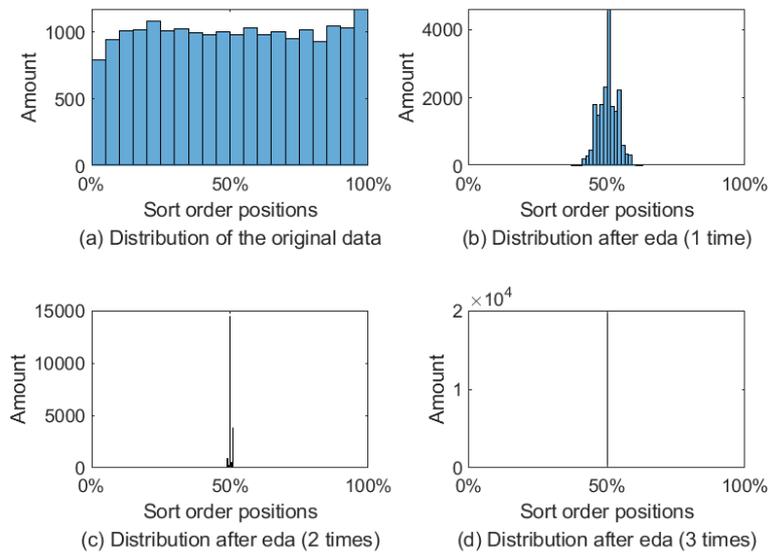

Fig. 1. Consistency of the order for uniform distribution (20k nodes, e<0.01)

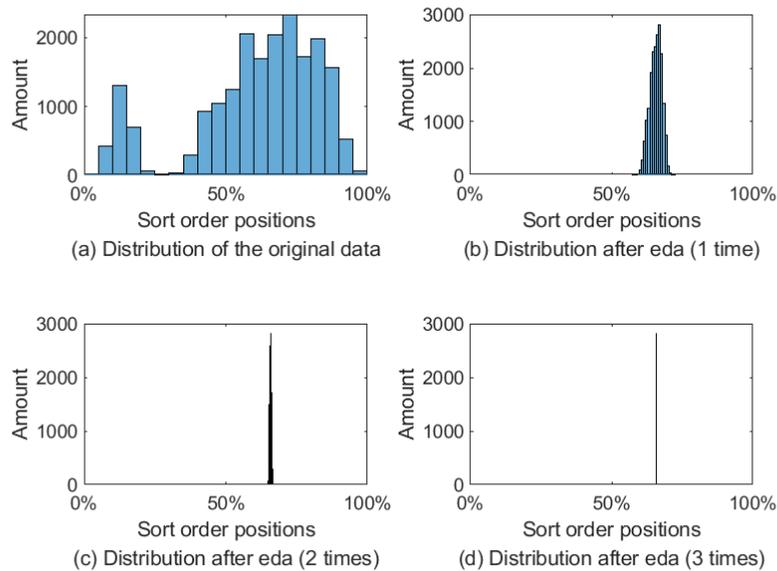

Fig. 2. Consistency of the order for random distribution (20k nodes, e<0.01)

## 4    Parallel processing

In the DIPS, multiple transactions always happened and need to be sorted simultaneously. What the EDA needs to do is to choose an appropriate function of $c(x)$ to make the result different for each transaction. For the fault tolerant range $\varepsilon<1\%$, the EDA can record up to 100 packages with multiple transactions contains. And if we set $\varepsilon<0.01\%$, up to 10000 packages will be accepted at the same time. The difference is just the number of rounds which the EDA needed for consensus. In an easy way to understand, the schematic workflow of parallel data sorting mechanism of the EDA can be separated into 4 parts as Fig. 3.

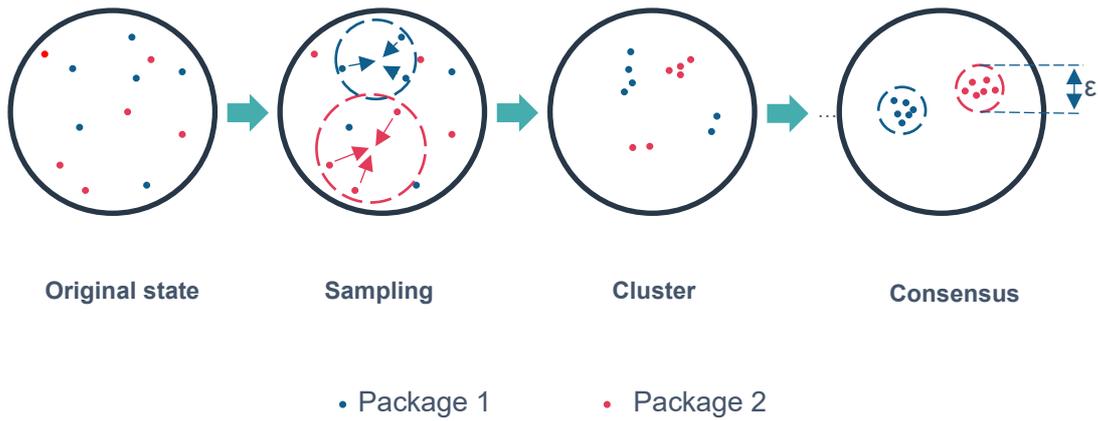

Fig. 3. The schematic workflow of the EDA in parallel consensus

We use 20k nodes to test the parallel capability of the EDA. As a preset condition, 1% peers are set as the failure node, which will do anything they want in the EDA network. The random minimum sample ratio is set as 1%, e<0.01. The result of each round can be seen in Fig. 4. In fact, the action of the fault nodes is random and worthless. In order to observe the process, we are closely displaying the situation of the fault nodes.

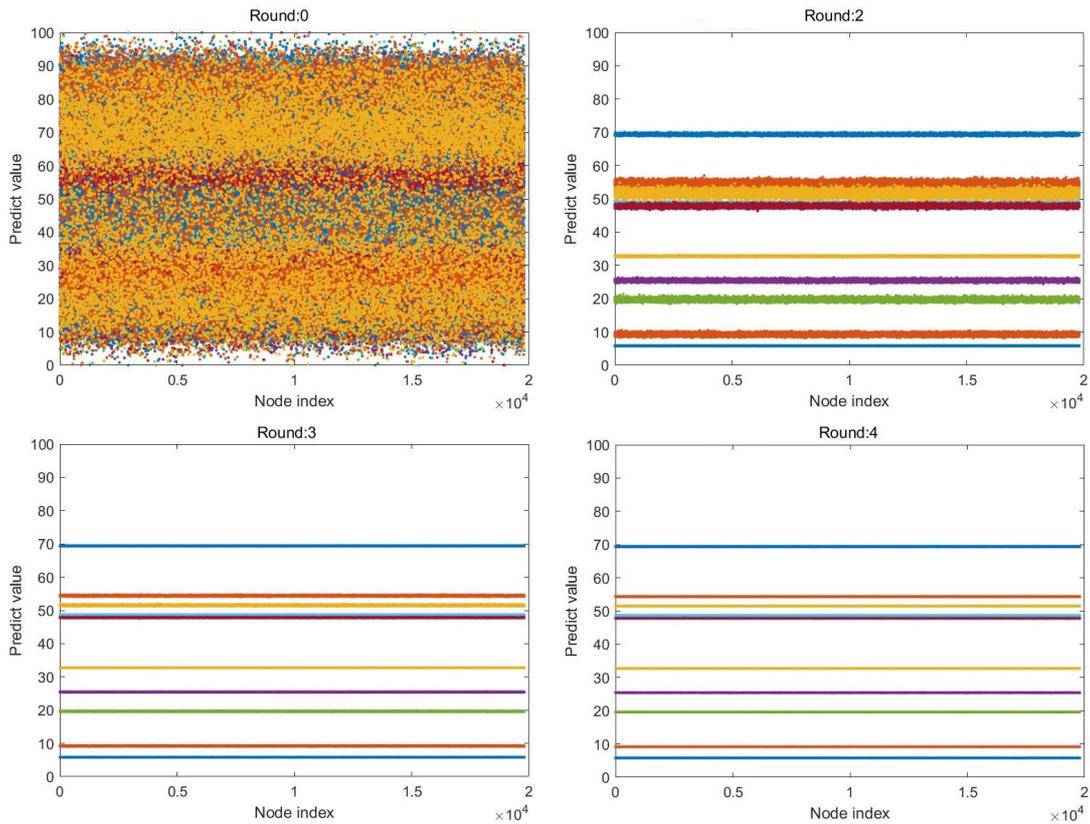

Fig. 4. The detail of parallel processing

# 5 Discussion

The EDA network is a totally different system. During system designing, a lot of interesting points were found. Even though the EDA exhibits some difficult problems, but the problems are resolved. We've realized that the EDA has a greater potential and application value. As a parallel sorting tool in large-scale cluster, the EDA could accelerate the consistency speed of the whole system, but the application scenario of the EDA is not limited to this.

**Real-random number generator**

The random number from random number generator is pseudo-random or real-random. The pseudo-random number generator generates the random number using embed time and other signal as seeds, while through physical processes in the true random number generator. EDA can be recognized as a real-random number generator, which is generated the random number by using sociology and cognitive behavior We call it real-random because of the uncontrollability of the total number, tendency and network environment of the system, which is different from the pseudo-random number generator.

**Distributed AI platform**

The workflow of EDA is trying to measure the difference through partial connections between nodes, tagging tags, and finally determine the characteristics (e.g. order, price) of the input through negotiation. It is the same way that how human recognize the world: everyone describes things according to his own way of observation and identifies the real content by way of negotiation. In this way, human evolved. So does the EDA: by neural network model and reward system standard, every participant will try to find out the system's optimal solution and win the prize. Also, through a large-scale network, the system will be become a full monte carlo measurement system, that will find out everything even we human don't know. It's not only a network, but a brain.